\documentclass{kapproc} 
\usepackage{procps} 
\usepackage[dvips]{graphicx}

\upperandlowercase
\kluwerbib

\begin{document}

\articletitle{Star Formation and Infrared Emission in Galaxies}

\author{Nikolaos D. Kylafis and Angelos Misiriotis}

\email{kylafis@physics.uoc.gr}

\affil{University of Crete\\
Department of Physics\\
P.O. Box 2208\\
71003 Heraklion, Crete\\
GREECE}

\begin{abstract}
The relationship between star formation and infrared emission in galaxies 
will be investigated.  If galaxies were simple objects and young stars were
completely covered with dust, then all the absorbed light of the young stars
would be re-emitted in the infrared and from the infrared emission of galaxies
we would infer the star formation rate (SFR) in them accurately.  To show the
complexities involved in real galaxies, we will use as a case study the
late-type spiral galaxies.  We will show that the heating of the dust is done
mainly by the UV radiation of the young stars and therefore the infrared
emission reveals the SFR in them.  With a realistic model and its application 
to a number of galaxies, tight correlations are derived between SFR and total
far infrared luminosity on one hand, and dust mass and 850 micron flux on the
other.  Other diagnostics of the SFR are examined and it is shown that there is
consistency among them.  Thus, the SFR for galaxies of all Hubble types has
been determined as well as for interacting starburst galaxies.  Combining
different methods, the star-formation history of the universe has been
determined and will be shown.  Finally, some early results from the Spitzer
Space Telescope will be presented.
\end{abstract}


\section*{Introduction}
We will start by asking the naive and rhetoric question of why should
star formation and infrared emission in galaxies be connected.

If galaxies were simple objects, consisting of young stars 
surrounded by optically thick dust, then practically all the
luminosity of the young stars would be absorbed by the dust
and it would be re-emitted in the infrared (IR) part of the spectrum.
Then, observation of the IR luminosity of a galaxy would give
us, with the use of an initial mass function, 
the star formation rate (SFR) in the galaxy.

However, real galaxies are more complicated objects for the following
reasons:  1) They consist mainly of old stars, which may also 
contribute to the heating of the dust.  2) The dust is
distributed throughout a galaxy and not around its stars.  
3) The optical depth of the dust in
a galaxy varies significantly with position in the galaxy and
direction.  Therefore, detailed modeling of the stars (young and old)
and the dust in a galaxy is needed before trustworthy conclusions are
drawn.

In what follows, we will treat late-type spiral galaxies as a case
study.  Then we will extend our discussion to all types of galaxies.

\section{Model for late-type spiral galaxies}

\subsection{Modeling of the optical light}

\begin{figure}[ht]
\vskip.2in
\resizebox{\hsize}{!}{\includegraphics{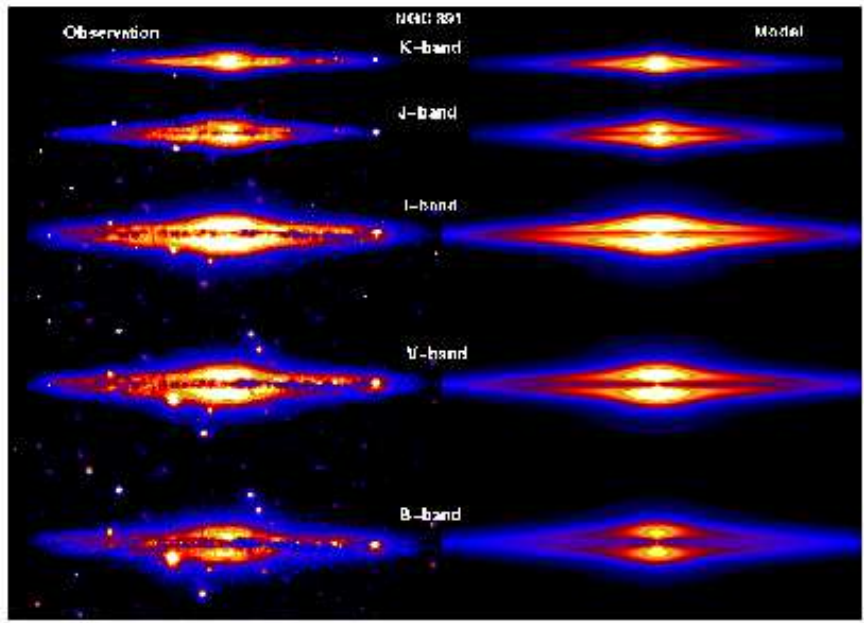}}
\caption{Images of NGC~891 in the K, J, I, V, and B bands and the corresponding
models.}
\label{fig:ngc891}
\end{figure}

\begin{figure}[ht]
\vskip.2in
\resizebox{\hsize}{!}{\includegraphics{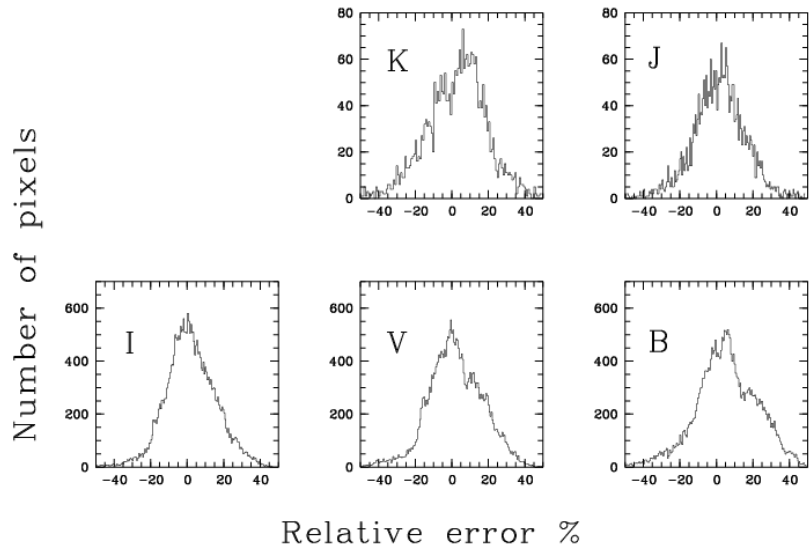}}
\caption{Histograms of the relative errors between the pixels
of the K, J, I, V, and B band observations and the corresponding models.}
\label{fig:histo}
\end{figure}

A successful model for late-type spiral galaxies has been applied to a number
of galaxies by Xilouris et al. (\cite{xilouris1999}).  
This model includes the old stellar
population in the form of an exponential (both in radius and height) disk and a
de Vaucouleurs (\cite{devaucouleurs1953}) spheroid.

The dust is distributed in another exponential
disk (with scales different than those of the stars).  Comparison of model
images with optical and near infrared (NIR) images of galaxies reveals the
total amount of dust (warm and cold) in them and its distribution.
Figure~\ref{fig:ngc891} shows NGC~891 in five bands and the corresponding model images 
(Xilouris et al. \cite{xilouris1998}). The agreement between the model and the observations
is very good. This is seen more clearly in Figure~\ref{fig:histo},
 where the residuals
between the model and the observations are shown 
(Xilouris et al. \cite{xilouris1998}).
The main conclusions of Xilouris et al. (\cite{xilouris1999}) are:

1) The scale height of the dust is about half that of the stars.

2) The radial scale length of the dust is about 1.4 times that of the stars.
Furthermore, the dust extends beyond the optical disk.

3) The average gas-to-dust ratio of seven spiral galaxies is about 400, i.e.
comparable to that of our Galaxy.

4) The extinction coefficient in the optical and NIR parts of the spectrum is
the same as in our Galaxy, indicating common dust properties among spiral
galaxies.

5) The central, face-on optical depth in the B-band of all seven galaxies is
less than one.  This means that, {\it if} all the dust has been accounted for
in these galaxies, then late-type spiral galaxies are transparent.

Similar conclusions have been reached by Alton et al. (\cite{alton1998}) and Davies et al.
(\cite{davies1999}).  The effects of spiral structure have been shown to be negligible
(Misiriotis et al. \cite{misiriotis2000}) and similarly for the effects of clumpiness in the
dust (Misiriotis and Bianchi \cite{misiriotis2002}).

\subsection{Modeling of the infrared emission}

\begin{figure}[ht]
\vskip.2in
\resizebox{\hsize}{!}{\includegraphics{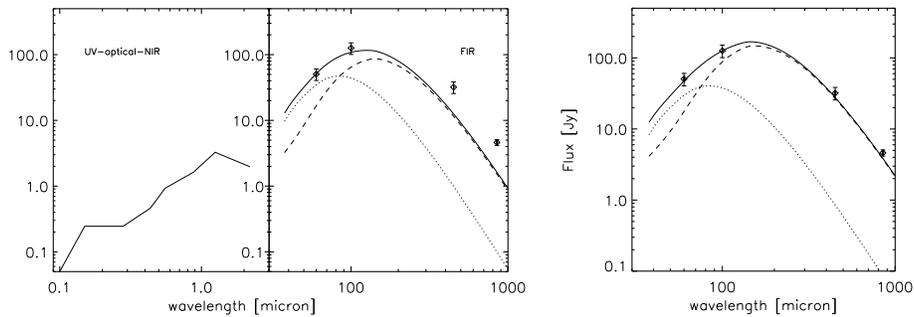}}
\caption{One- and two-dust-disk models of NGC~891.}
\label{fig:firmodel}
\end{figure}

In order to model the IR emission of late-type spiral galaxies, one has
to take into account the heating of the dust not only by the old stars but also
by the young ones.  This was done by Popescu et al.
(\cite{popescu2000}) for NGC 891 and by
Misiriotis et al. (\cite{misiriotis2001}) for four more galaxies.  
The model utilizes the dust distribution derived from the optical images and
assumes that the young stars (and therefore the UV emission)
are distributed in an exponential disk with a
small scale height (90 pc) and scale length equal to the B-band scale length
of the old stellar population. 
Part of the UV luminosity is absorbed by local sources (HII complexes) and the
rest is diffuse.
The far infrared (FIR) emissivity of the dust
was taken to be the same as the one thought appropriate for our Galaxy (Laor \&
Draine \cite{laor1993}).  

It was found that model fluxes in the submillimeter (submm) part of the
spectrum are significantly lower than the observed ones
(Fig. \ref{fig:firmodel}, left).  In order to account
for the ``missing'' submm flux, Popescu et al. (\cite{popescu2000}) 
proposed that there is
a second dust disk, not visible in the optical images, with scale height 90 pc
(equal to that of the young stars).  The mass in the second dust disk is about
the same as that of the first.  The model accounts 
(Fig. \ref{fig:firmodel}, right) not only for the spectral
energy distribution (SED) of NGC 891 in the FIR/submm regime (Popescu et al.
\cite{popescu2000}), but also for its observed surface 
brightness at 170 and 200 $\mu$m
(Popescu et al. \cite{popescu2004}).  
Furthermore, the model showed that the heating of the 
dust is done mainly by the UV.  Thus, the star formation rate in NGC 891
is directly connected to its infrared emission.

It is, however, possible that the FIR/submm dust emissivity value used for our
Galaxy (e.g. Draine \& Lee \cite{draine1984}) has been underestimated.  
Alton et al. (\cite{alton2004}),
using a simple but reliable model, studied the edge-on galaxies NGC 891, NGC
4013, and NGC 5907 and concluded that the emissivity at 850 $\mu$m is about
four times the widely adopted value of Draine \& Lee (\cite{draine1984}).
A detailed model by Dasyra et al. (\cite{dasyra2004}) arrived at
the same conclusion.  

At the moment, the above degeneracy (i.e., more dust or higher emissivity) is
not lifted, but the higher emissivity seems to be a real possibility
(Dasyra et al. \cite{dasyra2004}).

\section{Correlation between star formation and infrared emission}

\begin{figure}[ht]
\vskip.2in
\resizebox{\hsize}{!}{\includegraphics{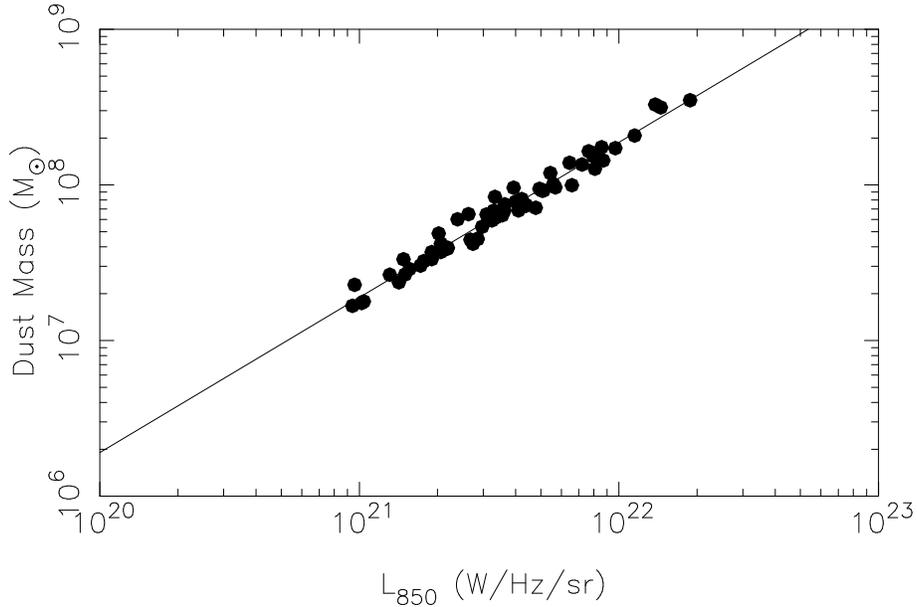}}
\caption{Dust mass as a function of luminosity at 850 $\mathrm{\mu m}$.
The solid line shows the best linear fit (in log-log space).}
\label{fig:coreldust}
\end{figure}

\begin{figure}[ht]
\vskip.2in
\resizebox{\hsize}{!}{\includegraphics{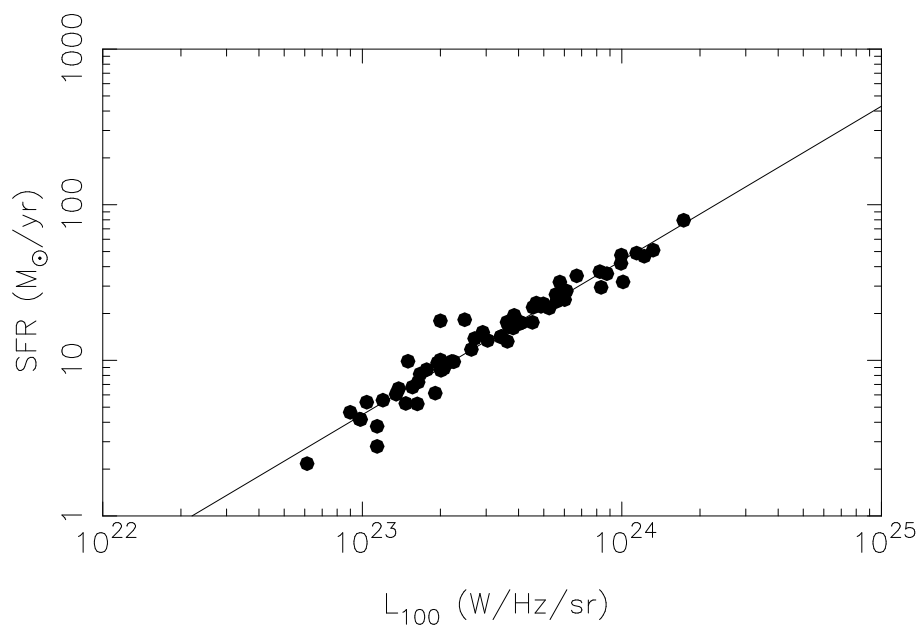}}
\caption{Star formation rate plotted as a function of $L_{100}$.
The solid line is the best linear fit (in log-log space).}
\label{fig:corel100}
\end{figure}

\begin{figure}[ht]
\vskip.2in
\resizebox{\hsize}{!}{\includegraphics{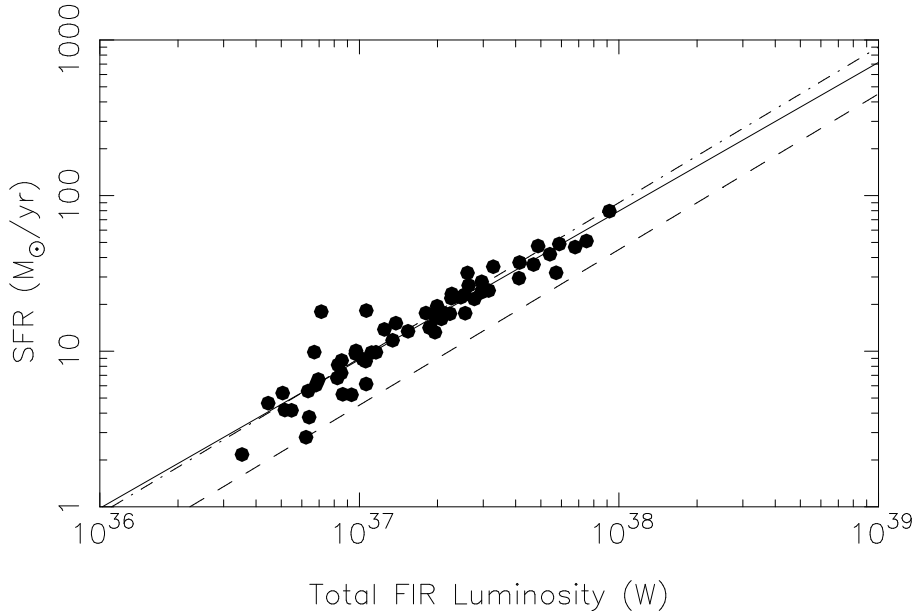}}
\caption{Star formation rate as a function of total FIR luminosity.
The solid line is the best linear fit (in log-log space).
The dashed line is drawn according to the SFR-$L_{FIR}$ relation of
Kennicutt (\cite{kennicutt1998}) for starburst galaxies. The dash-dotted line
is Kennicut's relation if one 50\% rather than 100\% transformation efficiency
of UV into FIR}
\label{fig:corelIR}
\end{figure}

As we saw above, in order for one to fit the SED of a late-type spiral galaxy,
a SFR must be assumed.  The question then is:  Can a
correlation be found between SFR and infrared emission for a sample of late-type
spiral galaxies?

The answer is yes and it was given by Misiriotis et al. (\cite{misiriotis2004}). 
They studied 62 bright IRAS galaxies from the SCUBA Local Universe Galaxy 
Survey of Dunne et al. (\cite{dunne2000}).  
Figure~\ref{fig:coreldust} shows the derived dust mass of the galaxies as a function of
their 850 $\mu$m luminosity.  The correlation is linear and quite impressive.
This is not surprising, because most of the dust in these galaxies 
(as in the galaxies studied by Xilouris et al. 1999) is cold ($\sim 15$ K).  
In Figure~\ref{fig:corel100}, the derived SFR is shown as a function of the 100 $\mu$m
luminosity.  Since the peak of the SED of these galaxies is near 100 $\mu$m, it
is again not surprising that the correlation is linear and quite tight.  
In Figure~\ref{fig:corelIR}, 
the sought after correlation between SFR and IR luminosity is shown.
The solid line is a linear fit to the data, while the dashed one is the
correlation of Kennicutt (\cite{kennicutt1998}; see also Buat \& Xu \cite{buat1996}).
The dot-dashed line shows the Kennicutt correlation if one assumes that the
UV to IR transformation efficiency is not 100\% (as in starbursts) but 50\%.

\section{Other star formation diagnostics}
Several other SFR diagnostics have been used over the years and the question is
how well they agree among themselves.  The answer now is quite well, though
this was not the case a few years ago.

Kewley et al. (2002) showed that the SFR derived from the IR luminosity and
that derived from ${\rm H\alpha}$ (properly corrected for extinction) are in
excellent agreement {\it for all types of galaxies} (from elliptical, to
spiral, to peculiar) and {\it for SFR ranging over four orders of magnitude}.
The reader is also referred to the work of Buat et al. (\cite{buat2002}), Hirashita et
al. (\cite{hirashita2003}), Panuzzo et al. (\cite{panuzzo2003}),
and Flores et al. (\cite{flores2004}).

The forbidden lines of [OII] have also been used as a measure of SFR.  Kewley
et al. (\cite{kewley2004}) showed that the SFR derived from [OII] and that derived from
${\rm H\alpha}$ are in excellent agreement for all types of galaxies and for four
orders of magnitude of the SFR.

Not surprisingly, the SFR determined from UV observations is in excellent
agreement with the SFR determined from the IR 
(Iglesias-P\'{a}ramo et al. \cite{iglesias-paramo2004}).
On the other hand, what is somewhat surprising is the fact that the 1.4 GHz
luminosity is a good indicator of the SFR.  
For an explanation of this, the reader is referred to Bressan et al. (\cite{bressan2002}),
but the explanation is not widely accepted 
(Bell \cite{bell2003}; Pierini et al. \cite{pierini2003}).
Afonso et al. (\cite{afonso2003}) showed that 
the SFR determined from the 1.4 GHz luminosity is in good agreement with the
SFR determined from ${\rm H\alpha}$ or [OII].

It is remarkable that even for interacting starburst galaxies there is a good
correlation between the SFR derived from the ${\rm H\alpha}$ flux and the SFR
derived from the FIR continuum (Dopita et al. \cite{dopita2002}).

Last but not least, we want to mention the work of F\"{o}rster Schreiber et al.
(\cite{forster2004}), who used a sample of galaxies containing a) quiescent spiral galaxies,
b) spiral galaxies with active circumnuclear regions, c) starburst galaxies, d)
LIRGs, and e) ULIRGs and showed that the monochromatic 15 $\mu$m continuum and
the 5 - 8.5 $\mu$m emission constitute excellent indicators of SFR.

We believe that the above are more than convincing that the determination of the
SFR in galaxies is a mature subject.  Significant progress has also been made
in the determination of the SFR history.  The most recent work in this subject
is that of Heavens et al. (\cite{heavens2004}), who
did an analysis of the `fossil record' of the current stellar populations of
96,545 nearby galaxies, from which they obtained a complete star-formation
history.  They found that the peak of star formation was at about five billion
years ago, i.e. more recently than other studies had found.  They also found
that the bigger the stellar mass of the galaxy, the earlier the stars were
formed, which indicates that high- and low-mass galaxies have very different
histories.

\section{Instead of a summary}
Instead of a summary, we want to mention just three of the recently reported
very exciting results from the Spitzer Space Telescope.

Appleton et al. (\cite{appleton2004}) reported that the FIR - radio correlation 
is valid to at least $z=1$ and similarly for the mid-infrared - radio correlation.

Higdon et al. (\cite{higdon2004}; see also Charmandaris et al. \cite{charmandaris2004}) 
reported that the redshift of high-$z$, faint galaxies can be 
determined from the 14 - 38 $\mu$m
spectrum!  This opens up tremendous possibilities for the determination of $z$
of faint galaxies.

Houck et al. (\cite{houck2004}) reported that the blue compact dwarf galaxy SBS 0335-052,
which has very low metallicity ($Z \sim Z_{\odot}/41$), has a featureless
mid-infrared spectrum with a peak at $\sim 28$ $\mu$m!  Taken at face value,
this means that there is no cold dust in this galaxy.

The reader is referred to the Special Issue of ApJS (Volume 154) for additional exciting
results from Spitzer.






%





\begin{chapthebibliography}{1}

\bibitem[2003]{afonso2003}
Afonso, J., Hopkins, A., Mobasher, B., Almeida, C.
2003, ApJ, 597, 269

\bibitem[1998]{alton1998}
Alton, P. B., Trewhella, M., Davies, J. I., et al.
1998, A\&A, 507, 125

\bibitem[2004]{alton2004}
Alton, P. B., Xilouris, E. M., Misiriotis, A., Dasyra, K. M., Dumke, M.
2004, A\&A, 425, 109

\bibitem[2004]{appleton2004}
Appleton, P. N., Fadda, D. T., Marleau, F. R., et al.
2004, ApJS, 154, 147

\bibitem[2003]{bell2003}
Bell, E. F.
2003, ApJ, 586, 794

\bibitem[2002]{bressan2002}
Bressan, A., Silva, L., Granato, G. L.
2002, A\&A, 392, 377

\bibitem[2002]{buat2002}
Buat, V., Boselli, A., Gavazzi, G., Bonfanti, C.,
2002, A\&A, 383, 801

\bibitem[1996]{buat1996}
Buat, V., Xu, C.
1996, A\&A, 306, 61

\bibitem[2004]{charmandaris2004}
Charmandaris, V.; Uchida, K. I.; Weedman, D., et al.,
2004, ApJS, 154, 142

\bibitem[2004]{dasyra2004}
Dasyra, K. M., Xilouris, E. M., Misiriotis, A., Kylafis, N. D.
2004,A\&A submitted 

\bibitem[1999]{davies1999}
Davies, J. I., Alton, P., Trewhella, M., Evans, R., Bianchi, S.
1999, MNRAS, 304, 495

\bibitem[1953]{devaucouleurs1953}
de Vaucouleurs, G.
1953, MNRAS, 113, 134

\bibitem[2002]{dopita2002}
Dopita, M. A., Pereira, M., Kewley, L. J., Capaccioli, M.
2002, ApJS, 143, 47

\bibitem[1984]{draine1984}
Draine, B. T., Lee, H. M.
1984, ApJ, 285, 89

\bibitem[2000]{dunne2000}
Dunne, L., Eales, S., Edmunds, M., Ivison, R., Alexander, P., Clements, D. L.
2000, MNRAS, 315, 115

\bibitem[2004]{flores2004}
Flores, H., Hammer, F., Elbaz, D., et al.
2004, A\&A, 415, 885

\bibitem[2004]{forster2004}
F\"{o}rster Schreiber, N. M., Roussel, H., Sauvage, M., Charmandaris, V.
2004, A\&A, 419, 501

\bibitem[2004]{heavens2004}
Heavens, A., Panter, B., Jimenez, R., Dunlop, J.
2004, Natur, 428, 625

\bibitem[2004]{higdon2004}
Higdon, S. J. U., Weedman, D., Higdon, J. L., et al.
2004, ApJS, 154, 174

\bibitem[2003]{hirashita2003}
Hirashita, H., Buat, V., Inoue, A. K.
2003, A\&A, 410, 83

\bibitem[2004]{houck2004}
Houck, J. R., Charmandaris, V., Brandl, B. R.
2004, ApJS, 154, 211

\bibitem[2004]{iglesias-paramo2004}
Iglesias-P\'{a}ramo, J., Buat, V., Donas, J., Boselli, A., Milliard, B.
2004, A\&A, 419, 109

\bibitem[1998]{kennicutt1998}
Kennicutt, R. C., Jr.
1998, ARA\&A, 36, 189

\bibitem[2002]{kewley2002}
Kewley, L. J., Geller, M. J., Jansen, R. A., Dopita, M. A.
2002AJ, 124, 3135

\bibitem[2004]{kewley2004}
Kewley, L. J., Geller, M. J., Jansen, R. A.
2004, AJ, 127, 2002

\bibitem[1993]{laor1993}
Laor, A., Draine, B. T.
1993, ApJ, 402, 441

\bibitem[2002]{misiriotis2002}
Misiriotis, A., Bianchi, S.
2002, A\&A, 384, 866

\bibitem[2000]{misiriotis2000}
Misiriotis, A., Kylafis, N. D., Papamastorakis, J., Xilouris, E. M.
2000, A\&A, 353, 117

\bibitem[2004]{misiriotis2004}
Misiriotis, A., Papadakis, I. E., Kylafis, N. D., Papamastorakis, J.
2004, A\&A, 417, 39

\bibitem[2001]{misiriotis2001}
Misiriotis, A., Popescu, C. C., Tuffs, R., Kylafis, N. D.
2001, A\&A, 372, 775

\bibitem[2003]{panuzzo2003}
Panuzzo, P., Bressan, A., Granato, G. L., Silva, L., Danese, L.
2003, A\&A, 409, 99

\bibitem[2003]{pierini2003}
Pierini, D., Popescu, C. C., Tuffs, R. J., V\"{o}lk, H. J.
2003, A\&A, 409, 907
              
\bibitem[2000]{popescu2000}
Popescu, C. C., Misiriotis, A., Kylafis, N. D., Tuffs, R. J., Fischera, J.
2000, A\&A, 362, 138

\bibitem[2004]{popescu2004}
Popescu, C. C., Tuffs, R. J., Kylafis, N. D., Madore, B. F.
 2004, A\&A, 414, 45

\bibitem[1999]{xilouris1999}
Xilouris, E. M., Byun, Y. I., Kylafis, N. D., Paleologou, E. V., Papamastorakis, J.
1999, A\&A, 344, 868

\bibitem[1998]{xilouris1998}
Xilouris, E. M., Alton, P. B., Davies, J. I.; Kylafis, N. D., Papamastorakis, J., Trewhella, M.
1998, A\&A, 331, 894

\end{chapthebibliography}
\end{document}